\documentclass[%
 reprint,
 amsmath,amssymb,
 aps,
]{revtex4-2}

\usepackage{graphicx}% Include figure files
\usepackage{dcolumn}% Align table columns on decimal point
\usepackage{bm}% bold math
\usepackage{physics}
\usepackage{underscore}
\usepackage{subcaption}
\usepackage{caption}
\usepackage{algorithm2e}
\usepackage{ragged2e}
\usepackage{algorithmic}
\usepackage{xcolor}
\newcommand{\harf}{\mbox{$\frac12$}}
\newcommand{\fr}{\xi}
\newcommand{\mr}{\mu}
\newcommand{\br}{\beta}
\newcommand{\Schr}{Schr\"{o}dinger }
\newcommand{\Vis}{\mathcal{V}}

\begin{document}

\preprint{APS/123-QED}

\title{\Schr cat states of a macroscopic charged particle co-trapped with an ion}

\author{S. Leontica}
\affiliation{Clarendon Laboratory, Parks Road, OX1 3PU}

\author{C. J. Foot}
\affiliation{Clarendon Laboratory, Parks Road, OX1 3PU}

\date{\today}% It is always \today, today,
             %  but any date may be explicitly specified

\begin{abstract}
We investigate the feasibility of observing matter-wave interference of a micron-sized charged particle by putting it into a quantum superposition of states with a distinguishable separation. In the proposed method, an atomic ion is confined in a linear Paul trap along with the massive charged particle so that we can make use of the extensive toolbox of experimental techniques developed to control quantum states of trapped ions, and to manipulate their motions with high fidelity operations. This approach provides a stringent test of the predictions of dynamical reduction models of delocalised quantum superpositions of a particle, reaching macroscopicities of up to $\mathcal{M}=17$.
\end{abstract}

%\keywords{Suggested keywords}%Use showkeys class option if keyword
                              %display desired
\maketitle

%\tableofcontents

% Questions for ion trappers:
% - What is the highest frequency field you can apply to modulate the spring constant of the trap, eg through the amplitude of the oscillating quadrupole field (helical resonator?)
% - What other modulations can be applied? eg tickle electrode
% - values for f and g - timescales?
% - Examples of modulation in the literature?
% - What is the macroscopic particle?
% - Measurement question
% - Does the protocol already exist? this hamiltonian - does it have a solution? defined as eqn 12, 29a,b. Solution non-rwa? (I think this probably relates to their nature paper with vera).

\section{Introduction}

Since the early days of quantum mechanics it has been debated whether there is an intrinsic limit to size of objects for which superposition of quantum states can be observed. Nowadays, about a century later, there are many powerful experimental techniques that can probe the quantum-classical cross-over and investigate phenomenological models of wave-function collapse that have been proposed such as continuous spontaneous localisation (CSL) and gravity-induced collapse \cite{RevModPhys2013}.  This paper describes a theoretical study of a novel approach that extends the experimental techniques that have been developed to perform high-fidelity operations with atomic ions in electrodynamic (Paul) traps, for work with charged particles of much higher mass. We envision a hybrid system with the two species of charged particles confined in close proximity such that the atomic ions, that interact with laser radiation, provide a `handle' for manipulating the quantum states of the micron-sized objects. The electrostatic force between the two different types of charged particle allows us to exploit sophisticated laser techniques that have been developed in the context of single-ion clocks and quantum information processing with linear arrays of trapped ions. Systems of two atomic ions with different masses have been studied in detail, for key applications in the transfer of quantum information between different species~\cite{Schmidt2005,PhysRevLett.125.080504, Wineland1998}. This allows the exchange of quantum information between an atomic species that has transitions amenable for excitation using laser radiation and other species that are otherwise inaccessible. In a recent breakthrough, coherent laser spectroscopy of highly charged ions was carried out by confining a Ar$^{+13}$ ion within a cloud of laser-cooled lighter ions ($^9$Be$^+$). Quantum logic was used to readout information about the highly charged ion via the fluorescence from the lighter ions \cite{Micke2020}. 

Numerous methods have been developed to observe matter-wave interferometry. Interference of the wavepackets of rubidium atoms with a spatial splitting of $0.5$\,m between the two arms of the interferometer has been demonstrated \cite{Kovachy2015}. Work to increase the mass of the particles undergoing interference has been carried out with large molecules of mass $2\times 10^{-23}$\,kg going through gratings with period 266\,nm \cite{Nimmrichter2013,Haslinger2013,Eibenberger2013,largestmacro}.
In recent years there have also been many experiments directed towards preparing oscillators in their ground state. This can be achieved by cryogenic techniques for solid-state systems such as beams or cantilevers that have high oscillation frequencies; it is more difficult for free particles  but a breakthrough has recently been achieved experimentally by cooling a trapped nanoparticle in an optical cavity \cite{glasssphere, Magrini2021}. There are also proposals to use levitated nano- and micro-diamonds that contain a single nitrogen vacancy (NV) centre and so these particles have an electronic spin degree of freedom \cite{Wan2016} that facilitates the formation of superpositions. These massive objects with spin have parallels with the system investigated here (and similar ideas are discussed in \cite{Martinetz2020}), namely the atomic ion has spin (which provides a handle on different quantum states) and the strong repulsive electrostatic interaction leads to collective oscillations of the two-particle system.

In this design study we show how the laser techniques that have been developed for atomic ions in Paul traps can be applied to indirectly control the motional degrees of freedom of more massive charged objects, e.g.\ masses of around $10^{9}$\,u ($1.7\times 10^{-18}$\,kg). This approach confines both types of charged particles by electrodynamic fields in ultra-high vacuum and has several desirable features. Although trapping such highly charged particles is not within current capabilities, there is experimental progress in that direction. 

Here we describe the principle and explore the physics underlying the measurements on the assumption that suitable starting conditions can be obtained experimentally. We present estimates to show the feasibility of reaching these conditions with advances in experimental techniques for loading micron-sized objects with a charge-to-mass ratio within a few orders of magnitude of that of high-mass atomic ions, e.g.\ Yb$^+$. %) , only some of which are explored here.
The motivation for pursuing this goal is two-fold: i) to probe quantum macroscopicity by putting objects into a superposition of quantum states \cite{Millen2020}, and verifying this by matter-wave interference, as well as ii) providing technology that can make sensitive quantum sensors.\\

\section{Vibrational mode analysis}

Recent developments show that it is possible to drive Paul traps with different frequencies in order to effectively confine charged objects of largely different charge to mass ratios \cite{twofreq}.
In our theoretical treatment we assume two pointlike particles of masses $m$, $M$ and charges $e$ and $Q$ are confined to the central axis of the Paul trap and the radial oscillation modes do not become excited during the experiment. As a result we will only retain the form of the potential created by the trap along its axis of symmetry. Taking this into account we can express the potential energy of the crystal as:

\begin{equation}
    V= \harf m \omega ^2 _1 z_1 ^2 + \harf M \omega ^2 _2 z^2 _2 + \frac{1}{4\pi \epsilon_0}\cdot \frac{eQ}{\abs{z_2-z_1}},
        % V= \frac{1}{2}m \omega ^2 _1 z_1 ^2 + \frac{1}{2}M \omega ^2 _2 z^2 _2 + \frac{1}{4\pi \epsilon_0}\cdot \frac{eQ}{z_2-z_1},
\end{equation}

where $\omega _1$ and $\omega _2$, the angular oscillation frequencies of the particles in the trap %, are related by $\omega_2 ^2 =\fr \omega_1 ^2 $, with
%$\fr = \frac{Q}{M}\cdot\frac{m}{e} \ll 1$.
%\begin{equation}\label{eq:QMra}
%    \fr = \frac{Q}{M}\cdot\frac{m}{e} < 1
%\end{equation} Note that 
and $z_1$ and $z_2$ are the absolute positions in the trap. Since the trapping along this axis is entirely electrostatic we can express $\omega_2$ in terms of $\omega_1$ as 

\begin{equation}
    \omega_2^2 = \frac{Q}{e} \omega_1^2.
\end{equation}

We assume $z_1<z_2$, so the ion is always to the left of the macroscopic particle. In this form the equilibrium positions of the two particles with respect to the center of the trap are given by:

\begin{subequations}
\begin{align}
    d^3 = \frac{eQ}{4\pi \epsilon_0 m \omega_1^2}&(1+\frac{e}{Q}) \approx \frac{eQ}{4\pi \epsilon_0 m \omega_1^2}, \\
    z^{(0)}_1 &= -\frac{Qd}{e+Q}, \\
    z^{(0)}_2 &= \frac{ed}{e+Q},
\end{align}
\end{subequations}

where $d$ is the distance between the particles at equilibrium and $z_1^{(0)},z_2^{(0)}$ are the positions with respect to the trap center at equilibrium.
We now transform from absolute positions to displacement variables for the particles and describe the motion of the two-ion crystal in terms of small oscillations around  equilibrium. This allows truncation of the series of the Coulomb potential after the second term thus transforming the problem into a system of two coupled oscillators. We discuss the validity region of this truncation in Appendix \ref{app:approximations}. This standard procedure results in new coordinates for the two normal modes, in-phase ($z_i$) and out-of-phase ($z_o$):

\begin{subequations}
    \begin{align}
        z_1 &= b_{1i}z_i + b_{1o}z_o, \\
        z_2 &= b_{2i}z_i + b_{2o}z_o. \label{eq:z2}
    \end{align}
\end{subequations}
%The parameters can be computed in terms of $\fr$ and $\mr = M/m \gg 1$ and the result is:
These are normalized such that

\begin{subequations}
\begin{align}
    z_i = \sqrt{\frac{\hbar}{2m\omega_i}}(a_i+a_i^{\dagger}), \label{eq:zi} \\
    z_o = \sqrt{\frac{\hbar}{2m\omega_o}}(a_o+a_o^{\dagger}).
\end{align}
\end{subequations}

\begin{figure}[t!]
	\centering{}
	\includegraphics[width=0.45\textwidth]{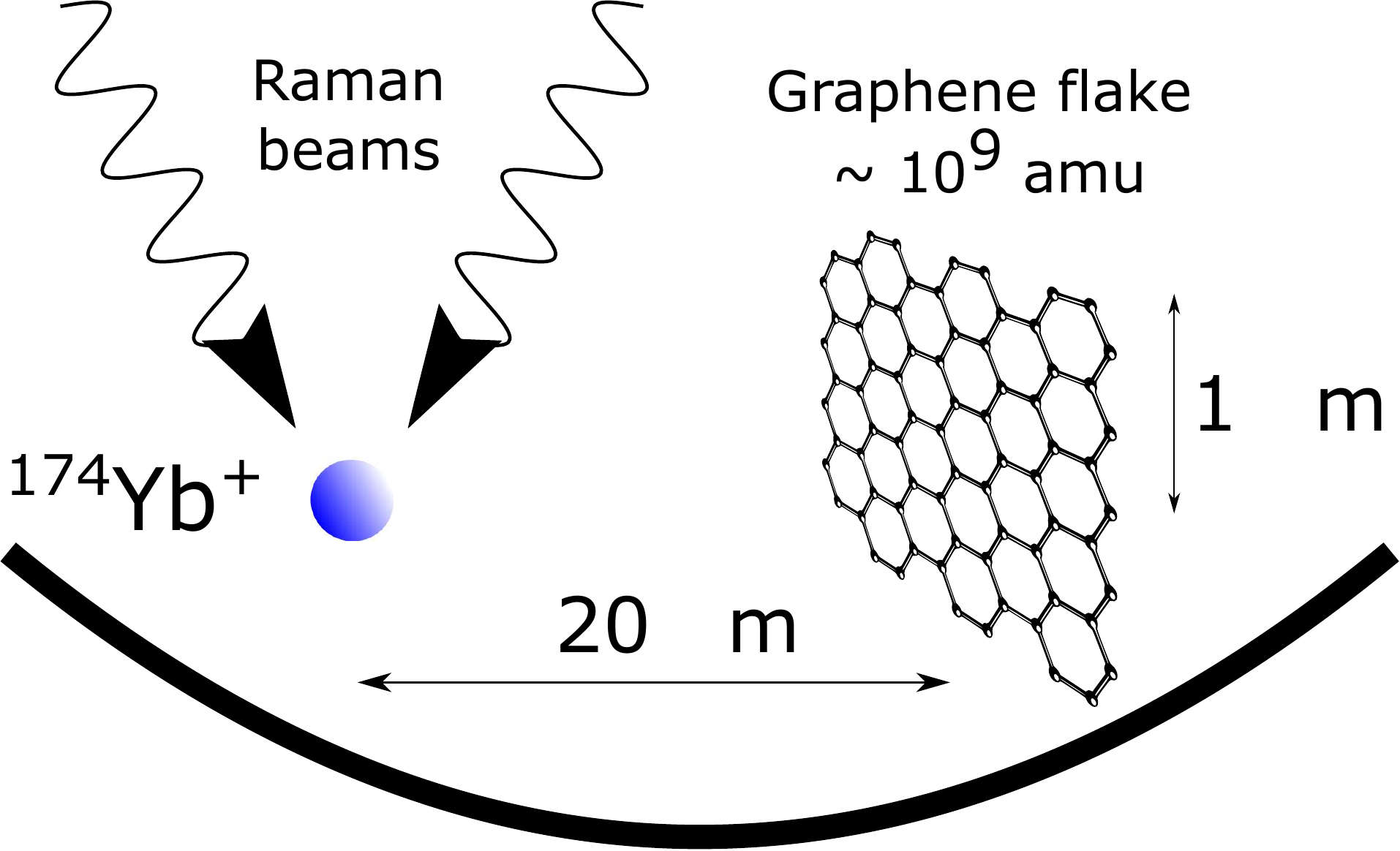}
	\caption{Schematic representation of the experimental setup. Ion and charged graphene flake are co-trapped in a longitudinal quadratic potential and interact through electrostatic repulsion. Raman beams are used to entangle the spin state of the ion to the common center-of-mass mode.}
	\label{fig:diagram}
\end{figure}

The amplitudes are determined by two parameters: the ratios of the masses $\mr$ and of the charge-to-mass ratios $\fr$,
    \begin{align}
        \mr &= \frac{M}{m}, \\
        \xi &=  \frac{Q}{M}\cdot \frac{m}{q} \label{eq:QMra}. %.
    \end{align}
For neatness we also define the quantity
    \begin{align}\label{eq:definebeta}
       \br &= \harf (3-\fr). 
    \end{align}
In terms of $\mr,\fr$ and $\br$ the mode amplitudes are  
% \begin{subequations}
%     \begin{align}
%         b_{1i} &= \frac{1}{\sqrt{1+\frac{\mr}{4}(3-\fr)^2}}, \\
%         b_{2i} &= b_{1i} \frac{3-\fr}{2}, \\
%         b_{2o} &= \frac{1}{\sqrt{\mr +\frac{\mr^2}{4}(3-\fr)^2}}, \\
%         b_{1o} &= -b_{2o}(3-\fr)\frac{\mr}{2}.
%     \end{align}
% \end{subequations}

\begin{subequations}
    \begin{align}
        b_{1i} &= \frac{1}{\sqrt{1+ \mr\,\br^2}}, \\
        b_{2i} &= \br\, b_{1i} ,\label{eq:b2ib1i} \\
%        b_{2o} &= \frac{1}{\sqrt{ \mr\left(1 + \mr\,\br^2\right) }}, \\
        b_{2o} &= \frac{1}{\sqrt{\mr}}\cdot  b_{1i}, \\
        b_{1o} &= -\mr\br\, b_{2o}. \label{eq:b1obo0} % \\        \br &= \harf (3-\fr). \nonumber
    \end{align}
\end{subequations}
For our case of interest, $\mr \gg 1$ and $\xi \ll 1$. Therefore $\br \simeq 3/2$ (in \ref{eq:definebeta}) so that in the in-phase mode the two particles have comparable amplitudes \eqref{eq:b2ib1i}.

The Hamiltonian can be expressed as the sum of two quantized harmonic oscillators written in terms of creation and annihilation operators defined the usual way:

\begin{equation}
    \hat{H} = \hbar \omega_i (\hat{a}^\dagger_i \hat{a}_i + \harf ) + \hbar \omega_o (\hat{a}^\dagger_o \hat{a}_o + \harf),
\end{equation}

where the angular frequencies of the decoupled oscillations are expressed as $\omega_i = \sqrt{\fr}\omega_1$ and $\omega_o = \sqrt{3}\,\omega_1$.

\section{Macroscopicity}

Choosing to work with a mass of around $M=10^9$\,u rules out a wider range of possible modifications of the Schr\"{o}dinger equation, in the sense described in \cite{Macroscopicity}. For the macroscopic particle, we assume a graphene flake of radius $R \approx 0.8\,\mu$m. This is favorable because it allows us to produce spatial superpositions larger than the thickness of the flake. It is desirable to work with an atomic ion that has a relatively low charge-to-mass ratio so in the rest of this paper we choose parameters for $^{174} \mathrm{Yb}^{+}$; a relatively massive atomic ion in common use for ion-trapping experiments. This sets the value of the mass quotient to $\mr = 5.7\times 10^6$. If we want a specific relationship between the center-of-mass and breathing mode frequencies then this also sets the value of the necessary charge. A ratio of $\omega_o/\omega_i = 200$, or $\fr =7.5 \times 10^{-5}$ in equation \eqref{eq:QMra}, implies the flake must be charged to a value of around $Q = 430 e$. The feasibility of obtaining this charge is discussed in Appendix \ref{app:charging}. We set the trap frequency to $\omega_1 / 2\pi = 1 \operatorname{MHz}$. This gives a distance between the atomic ion and the flake of $d=21\,\mu$m. For this choice of parameters we can estimate the macroscopicity from the formula \cite{Macroscopicity}:

\begin{equation}\mathcal{M} =\log _{10}\left[\left|\frac{1}{\ln f}\right|\left(\frac{M}{m_{e}}\right)^{2} \left(\frac{\Delta x}{R}\right)^2 \frac{t}{1 \mathrm{s}}\right],\end{equation}

where $\Delta x \approx 6$ \AA \space is the proposed size of the superposition and $R$ is the radius of the flake. If the decoherence rate can be made sufficiently low that spin coherence is retained, and interference fringes are observed, after the splitting time of $ \sim 1 \operatorname{ms}$ then we can obtain values in the range of $\mathcal{M} \approx 17$, three orders of magnitude higher than in previous work \cite{largestmacro}. The assumptions that lead to this estimate are elaborated in section \ref{simulation}.\\

We can also aim to study the behaviour of the system in the presence of spontaneous collapse type modifications of the Schr\"{o}dinger equation. For this, we assume the existence of an extra dynamical term in the von Neumann equation used to evolve the density matrix $\rho$:

\begin{subequations}
\begin{align}
&\frac{\partial \rho}{\partial t}=-\frac{i}{\hbar}[H, \rho] + \mathcal{L} \rho, \\
\label{modification}
\mathcal{L} \rho=\frac{1}{\tau}&\left[\int d^2\alpha g(\alpha) D(\alpha) \rho D^{\dagger}(\alpha)-\rho \right],
\end{align}
\end{subequations}

where the operators $D(\alpha)$ are phase space displacement operators, defined in \eqref{displacement}, $g(\alpha)$ is some phase space distribution and $\tau$ is a free parameter quantifying the frequency of the kicks. A theorem by Holevo \cite{HOLEVO1993211} proves that this is the most general modification which still preserves Galilean invariance. We follow \cite{Macroscopicity} and assume that the distribution $g(\alpha)$, which sets the probability of displacements of a given magnitude, is completely characterized by the standard deviations along the $x$ and $p$ axes of the phase space. From the effect of scaling equation \eqref{modification} from elementary particles to rigid composite systems we can speculate further that space translations should have negligible effect, leaving us with a single parameter $\sigma_p$ that specifies the distribution of kicks. This can be equivalently stated as $\sigma = \hbar/\sigma_p$, a length scale for the separation in a spatial superposition called the critical length. The strength of the modification lies in the coherence time parameter $\tau$, which depends on the mass of the macroscopic object as $\tau \propto m^{-2}$ in the region where the size of the object is much smaller than the critical length \cite{Macroscopicity}.  We define $\tau_e$, the coherence time of an electron, as a free parameter of the theory. %We expect 
Possible methods of distinguishing spontaneous collapses from other sources of noise that may be present during experimental operation include measuring the dependence on mass, shape and critical length. 

\section{Cooling}
%While still in its early stages, s
Several schemes have been proposed for cooling massive particles to their motional ground state \cite{Millenn2020,gonzalez}. Electrical cooling techniques based on coupling between the electrodes of the ion trap and outside RLC circuits have been studied theoretically \cite{resistivecooling}. Other researchers proved that it is possible to cool a glass sphere to its ground state using optical tweezers \cite{glasssphere}. These indicate that future technological advancement will enable the preliminary step of initialising a massive particle in its ground state of motion for levitated electromechanics experiments such as that proposed here. We show that our method is resilient to small initial excitation of the center-of-mass mode in a first-order approximation and therefore only requires the massive particle to be prepared in a low-lying thermal state rather than purely in the ground state. Since the out-of-phase mode is dominated by the atomic ion as seen in \eqref{eq:b1obo0}, we expect that cooling this mode to the ground state is equivalent to cooling the isolated ion and can be performed in the same way. Also the internal degrees of freedom of the macroscopic particle need not be cooled, as the use of ancillary ions avoids having to interact directly with the massive object. For the 2D graphene flakes considered in this work it is also necessary to cool the rotational degrees of freedom and obtain stable orientations. In principle, this would also allow the study of regimes of quantum mechanical behaviour where the particle undergoing interferometry is hot \cite{hotinterf}. The ability to hold the flake in different orientations may also enable the anisotropy of collapses for non-spherical objects to be used as a discriminator between CSL effects and heating from other sources, although we do not investigate this possibility in this paper.

\section{Theory}

\subsection{Spin-dependent force}

In order to create the desired spatial separation we consider two internal states of the ion, which we denote $\ket{\uparrow}$ and $\ket{\downarrow}$. To create Schr\"{o}dinger Cat states with the macroscopic particle we apply a state-dependent force resonant to the transition between the different energy states of the center-of-mass mode. By introducing a phase difference of $\pi$ between the forces driving the different spin states of the ion we can effectively create spatial superpositions of the common center-of-mass mode. This in turn delocalizes the massive flake without interacting with its internal degrees of freedom. In this paper, we consider the effect of  laser interactions driving Raman transitions to realise the initial splitting, but several other options exist. We assume a system of lasers which drive transitions between our spin states and another auxiliary state, as described in \cite{home2006entanglement}; the only part of the Hamiltonian that survives in the interaction picture takes the form

\begin{equation}
\label{hamiltI}
    \hat{H}_I = \sum_{m=\uparrow,\downarrow} \ket{m}\bra{m} \frac{\hbar \Omega_R}{2} e^{i(\delta kz_1-\delta\,t+\phi_m)} + h.a.
\end{equation}

where $\Omega_R$ is the Raman frequency, $\delta k$ is the wavenumber of the Raman pair and $\delta$ is the (angular frequency) detuning from resonance.
To shorten notation we will retain only one of the terms and drop the subscript $m$, keeping in mind throughout that the expression is true for both spin states. We now focus our attention on the spatial variation of the exponential in (\ref{hamiltI}) and introduce an expression for the position of the ion

\begin{equation}
\begin{split}
    e^{i\delta kz_1} =& e^{i\eta_i(\hat{a}^\dagger_i+\hat{a}_i)}e^{i\eta_o(\hat{a}^\dagger_o+\hat{a}_o)} \\
    =&(1+i\eta_i(\hat{a}^\dagger_i+\hat{a}_i)+\dots) \\ \times&(1+i\eta_o(\hat{a}^\dagger_o+\hat{a}_o)-\frac{1}{2}\eta_o^2(\hat{a}^\dagger_o+\hat{a}_o)^2+\dots),
\end{split}
\end{equation}

where $\eta_i$ and $\eta_o$ are the Lamb-Dicke parameters of the two modes \cite{home2006entanglement}. Assessing the importance of terms that appear in the expansion is not an easy task but we may ignore some terms based on two criteria: the Rabi frequency of the term is small compared to the detuning from resonance and the timescale of the evolution is long compared to the duration of the experiment. We expect the duration of the experiment to be on the scale of $\Delta t \approx \frac{1}{\Omega_R \eta_i} \approx 1\operatorname{ms}$, roughly equal to the inverse strength of the in-phase mode term in the expansion. Only terms that create variations on this timescale need to be considered, namely:

\begin{equation}
\begin{split}
    e^{i\delta kz_1} &= 1+i\eta_i(\hat{a}^\dagger_i+\hat{a}_i)+i\eta_o(\hat{a}^\dagger_o+\hat{a}_o) \\ 
    &-\frac{1}{2}\eta_o^2(\hat{a}^\dagger_o+\hat{a}_o)^2+\dots
\end{split}
\end{equation}

We analyse these three terms separately to understand their effect on the system. When the first term is driven on resonance we get a linearly increasing displacement of the in-phase mode (in opposite directions for the two spins) which we will refer to as the Spin-Dependent Force (SDF). The second and third terms excite the parasitic out-of-phase mode and it is desirable to drive this mode around closed loops in phase space during the experimental sequence to ensure maximum visibility of the interferometric fringes, as we shall show. This would be straightforward in the absence of the third term (corresponding to a second-order effect), which causes a slow but constant drift of the mode away from the origin of the phase space. We show that it is possible to treat all terms analytically and so design efficient simulations. The small Lamb-Dicke parameter of the in-phase mode (i.e. $2\times 10^{-4}$) allows us to neglect off-resonant coupling between the two modes, effectively splitting the problem into two separate quantum systems which evolve according to the interaction Hamiltonians

\begin{subequations}
\begin{align}
    %\hat{H}_{l} &= \hbar \Omega_R \cos(\delta t -\phi), \\
    \label{hami}
    \hat{H}_i &=  \hbar \Omega_R \eta_i \sin(\delta\, t -\phi)(a_i e^{i\omega_i t}+ a^\dagger_i e^{-i\omega_i t}), \\
    \begin{split}
    \hat{H}_o &= \hbar \Omega_R \eta_o \sin(\delta\, t -\phi)(a_o e^{i\omega_o t}+ a^\dagger_o e^{-i\omega_o t}) \\
    &- \frac{1}{2} \hbar \Omega_R \eta_o^2 \cos (\delta\, t -\phi)(a_o e^{i\omega_o t}+ a^{\dagger}_o e^{-i\omega_o t})^2.
    \end{split}
    \label{hamo}
\end{align}
\end{subequations}

\subsection{Parametric amplification}

By applying an RF potential to the electrode caps of the trap we can modulate the strength of harmonic oscillator and resonantly excite oscillation of the in-phase mode. This induces an exponential amplification of the amplitude if the frequency and phase are sufficiently steady over time, which we call Parametric Amplification (PA). Additionally, this approach helps to speed up the execution time and therefore reduce the accumulated decoherence associated with laser scattering. As the magnitude of all terms added to the Hamiltonian by this interaction is the same for all modes we will only consider the one that is resonant with the in-phase mode. This can be cast in the form

\begin{equation}
\label{hamiltPA}
    \hat{H}_i =  \harf\hbar \Omega_{PA} \sin(2 \omega_i t -\phi_{PA})(a_i e^{i\omega_i t}+ a^\dagger_i e^{-i\omega_i t})^2,
\end{equation}

with $\Omega_{PA}$ an interaction strength and $\phi_{PA}$ the phase of the modulation. The use of such terms has been considered in the context of quantum computing with trapped ions to speed up entangling gates \cite{PA1,PA2}; these references also provide a thorough analysis of the limiting factors of the approach and discuss various operation sequences which optimize the outcome fidelity.\\

There is only a single off-resonant term which is non-negligible which arises from the oscillating force acting on the out-of-phase mode. This additional Hamiltonian is 

\begin{equation}
    H_o^{(1)} = \hbar\Omega_{R}\eta_o^{(1)}\sin(2\omega_i t -\phi_{PA})(a_o e^{i\omega_o t}+ a^\dagger_o e^{-i\omega_o t}),
    \label{hamPAoff}
\end{equation}

which we express in the same form as the force from the SDF for convenience. The effective Lamb-Dicke parameter is given by 

\begin{equation}
    \eta_o^{(1)} = \frac{\Omega_{PA}}{\Omega_R}  b_{1o} z_1^{(0)} \sqrt{\frac{2m\omega_i^2}{\hbar \omega_o}}.
\end{equation}

If the driving intensity $\Omega_{PA}$ is large this term dominates the one originating from the SDF. There is a simple physical intuition behind this term and its dependence on the relevant parameters: when the strength of the longitudinal voltage is varied, the electric field produced by the endcap electrodes varies significantly at the position of the atomic ion, since its equilibrium position is displaced from the trap center. This gives a contribution to the potential energy proportional to the displacement from equilibrium, or equivalently a homogeneous oscillating force. More information about the approximations employed can be found in Appendix \ref{app:approximations}.

\subsection{Forced Harmonic Oscillator with modulated strength}

In order to learn more about the behaviour of the system let us move our viewpoint to a more general form of Hamiltonian  

\begin{equation}
\begin{split}
\label{hamiltonian}
    \hat{H} &= \hbar f(t)(a e^{i\omega t}+ a^\dagger e^{-i\omega t}) \\
    &+ \frac{1}{2}\hbar g(t)(a e^{i\omega t} + a^{\dagger} e^{-i\omega t})^2.
\end{split}
\end{equation}

This is a generalisation of the particular types of time-dependent Hamiltonians encountered in both the motion of the in-phase mode as given by equations \eqref{hami} and \eqref{hamiltPA}, as well as the out-of-phase mode driven by equation \eqref{hamo}.
A formal solution for the unitary evolution produced by this Hamiltonian is possible using a time-ordered exponential, but we will concentrate on a more practical class of analytical solutions expressible in terms of squeezed coherent states. The ability to solve the evolution analytically provides a powerful tool for the efficient description and simulation of the splitting protocol.

We introduce the displacement and squeeze operators which lie at the core of the analysis:

\begin{subequations}
        \begin{align}
    D(\alpha)&=\exp \left(\alpha \hat{a}^{\dagger}-\alpha^{*} \hat{a}\right), \label{displacement} \\
    S(\zeta)&=\exp \left[\frac{1}{2}\left(\zeta^{*} \hat{a}^{2}-\zeta \hat{a}^{\dagger 2}\right)\right].\label{squeeze}
    \end{align}
\end{subequations}

In terms of these, the squeezed coherent states are:

\begin{equation}\ket{\alpha, \zeta}=D(\alpha) S(\zeta)\ket{0}.\end{equation}

The question we want to answer is how the parameters $\alpha, \zeta$ evolve for this state for a Hamiltonian of the form \eqref{hamiltonian}. Since these states form an over-complete basis we can use P representations \cite{Prepresentation} to compute time evolution of arbitrary states and compute expectation values by integrals of Gaussian functions. The composition relations of the displacement and squeeze operators that are used in the calculations are 
\begin{subequations}
    \begin{align}
        \label{combD}
        \hat{D}(\alpha) \hat{D}(\beta)=&e^{\left(\alpha \beta^{*}-\alpha^{*} \beta\right) / 2} \hat{D}(\alpha+\beta), \\
        \begin{split}
        \label{combS}
            S\left(\zeta_{1}\right) S\left(\zeta_{2}\right)=&S\left(\zeta_{3}\right) \exp \left[\ln \frac{1+\tau_{1} \tau_{2}^{*}}{1+\tau_{1}^{*} \tau_{2}} \frac{\sigma_{3}}{2}\right], \\ \tau_{3}:=\frac{\tau_{1}+\tau_{2}}{1+\tau_{1}^{*} \tau_{2}}&, \quad \quad \sigma_{3}:=a^{\dagger} a+\frac{1}{2},
        \end{split}
    \end{align}
\end{subequations}

where $\tau_{i}$ is related to $\zeta_{i}$ for $i = 1,2,3$ by the relations

\begin{subequations}
\label{tauzeta}
\begin{align}
    \tau_i &= e^{i \theta_i} \tanh r_i, \\
    \zeta_i &= r_i e^{i \theta_i}.
\end{align}
\end{subequations}

We also use of the commutation relation between squeezing and displacement operators given by

\begin{equation}
    \begin{split}
    \label{comm}
        \hat{D}(\alpha) \hat{S}(\zeta)&=\hat{S}(\zeta) \hat{D}(\gamma), \\
        \text{where} \quad \alpha&=\gamma \cosh r-\gamma^{*} e^{i \theta} \sinh r.
        \end{split}
\end{equation}

We separate the full Hamiltonian into three parts whose action is more easily tractable in the algebraic manipulations:

\begin{subequations}
\begin{align}
    H_0 &= \hbar f(t)\left(ae^{i\omega t}+a^{\dagger}e^{-i\omega t}\right) \label{eq:H0ft}, \\
    H_1 &= \frac{1}{2}\hbar g(t)\left(a^2e^{2i\omega t}+a^{\dagger\hspace{1pt}2}e^{-2i\omega t}\right) \label{eq:H1gt},\\
    H_2 &= \hbar g(t)\left(a^{\dagger}a+\frac{1}{2}\right) = g(t)\sigma_3 \label{eq:H2gt}.
\end{align}
\end{subequations}

We begin by writing the infinitesimal time propagator of the system:

\begin{equation}
\begin{split}
    U&(t+dt,t) = \operatorname{I}- \frac{i}{\hbar}Hdt +\mathcal{O}(dt^2) \\
        &= e^{-\frac{i}{\hbar} H_0 dt}e^{-\frac{i}{\hbar} H_1 dt}e^{-\frac{i}{\hbar} H_2 dt}+\mathcal{O}(dt^2). \\
\end{split}
\end{equation}

Since the three exponentials are infinitesimally close to the identity we can study their effects separately on a squeezed coherent state and conjoin them at the end. The first \eqref{eq:H0ft} is simply an infinitesimal displacement operator of amplitude:

\begin{equation}
    d\alpha_1 = -if(t)e^{i\omega t}dt.
\end{equation}

This can be merged with the existing displacement $\alpha$, also giving rise to the well-known geometric phase:

\begin{equation}\label{eq:dPhi}
\begin{split}
    d\Phi_1 &= -\frac{f(t)}{2}\left(\alpha^\ast e^{i\omega t}+\alpha e^{-i\omega t}\right)dt \\
    &= -\frac{i}{2}\left(\alpha^{\ast}d\alpha_1-\alpha d\alpha_1^{\ast}\right).
    \end{split}
\end{equation}

The exponential of $H_1$ in \eqref{eq:H1gt} is a squeeze operator of magnitude:

\begin{equation}
    \delta z_1 = ig(t)e^{i2\omega t}dt.
\end{equation}

Using the commutation relation \eqref{comm} to interchange position with the displacement operator gives rise to an additional infinitesimal displacement given by:

\begin{equation}
        d\alpha_2 = -\alpha^{\ast}\delta z_1.
\end{equation}

This is proportional to the magnitude of $\alpha$ and is the primary term driving exponential amplification. It is worth noting that this appears directly inside the displacement operator, so it does not give rise to a phase similar to $d\Phi_1$ in \eqref{eq:dPhi}. The infinitesimal squeeze can now be combined with the squeezing operator using equation \eqref{combS}. The exponential acts directly on vacuum so we get $\sigma_3 = 1/2$ and it becomes a simple phase:

\begin{equation}
    d\Phi_2 = \frac{1}{4i}\left(\delta z_1 \tau^{\ast}-\delta z^{\ast}_1 \tau \right).
\end{equation}

Here $\tau$ is related to $\zeta$ by the definition in \eqref{tauzeta}. The change in $\tau$ can be expressed as:

\begin{equation}
    d\tau_1 = \delta z_1 - \tau^2 \delta z^{\ast}_1,
\end{equation}

\begin{figure*}
     \centering
     \begin{subfigure}[t]{0.3\textwidth}
         \centering
         \includegraphics[width=\textwidth]{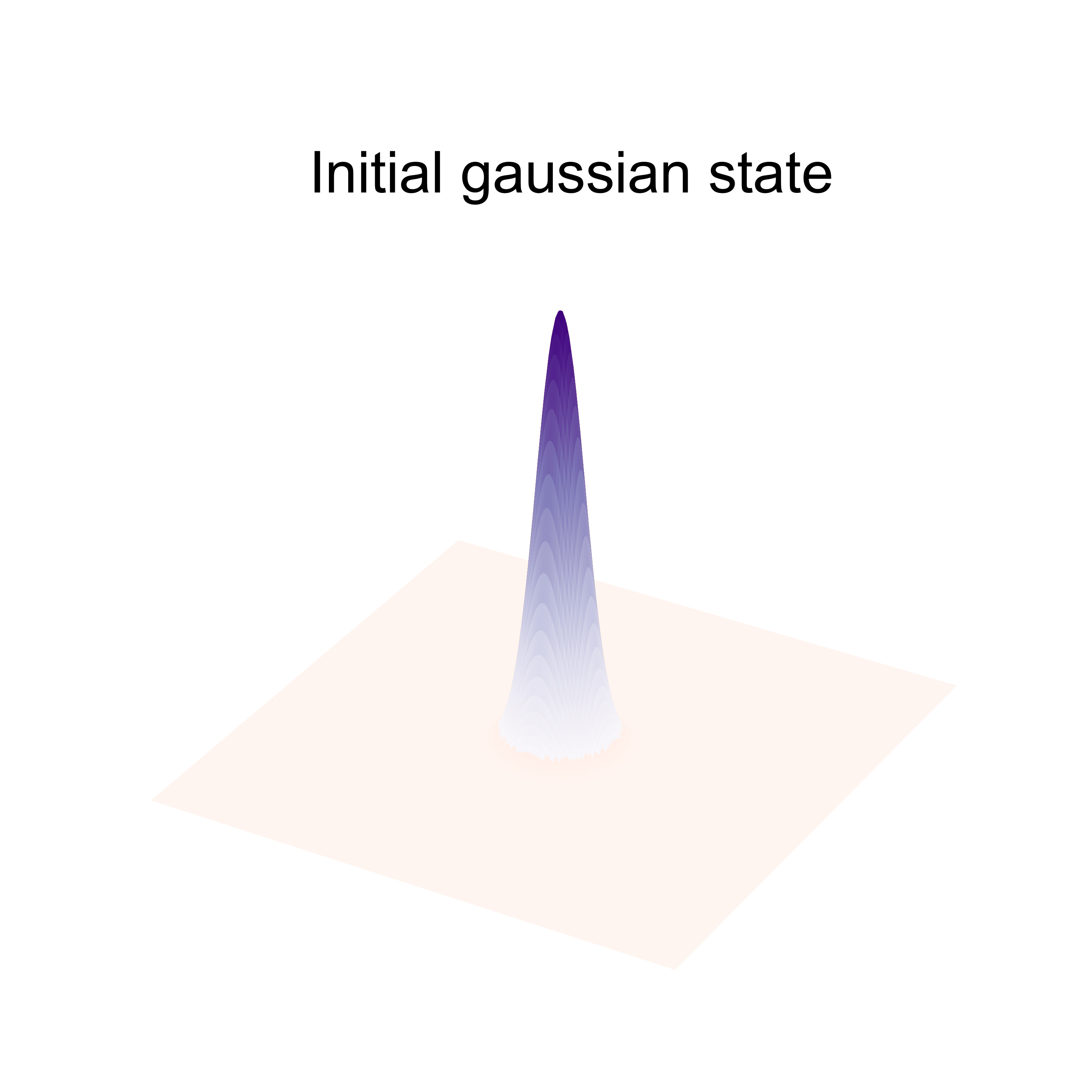}
         \label{fig:wigner1}
         \caption{}
     \end{subfigure}
     \hfill
     \begin{subfigure}[t]{0.3\textwidth}
         \centering
         \includegraphics[width=\textwidth]{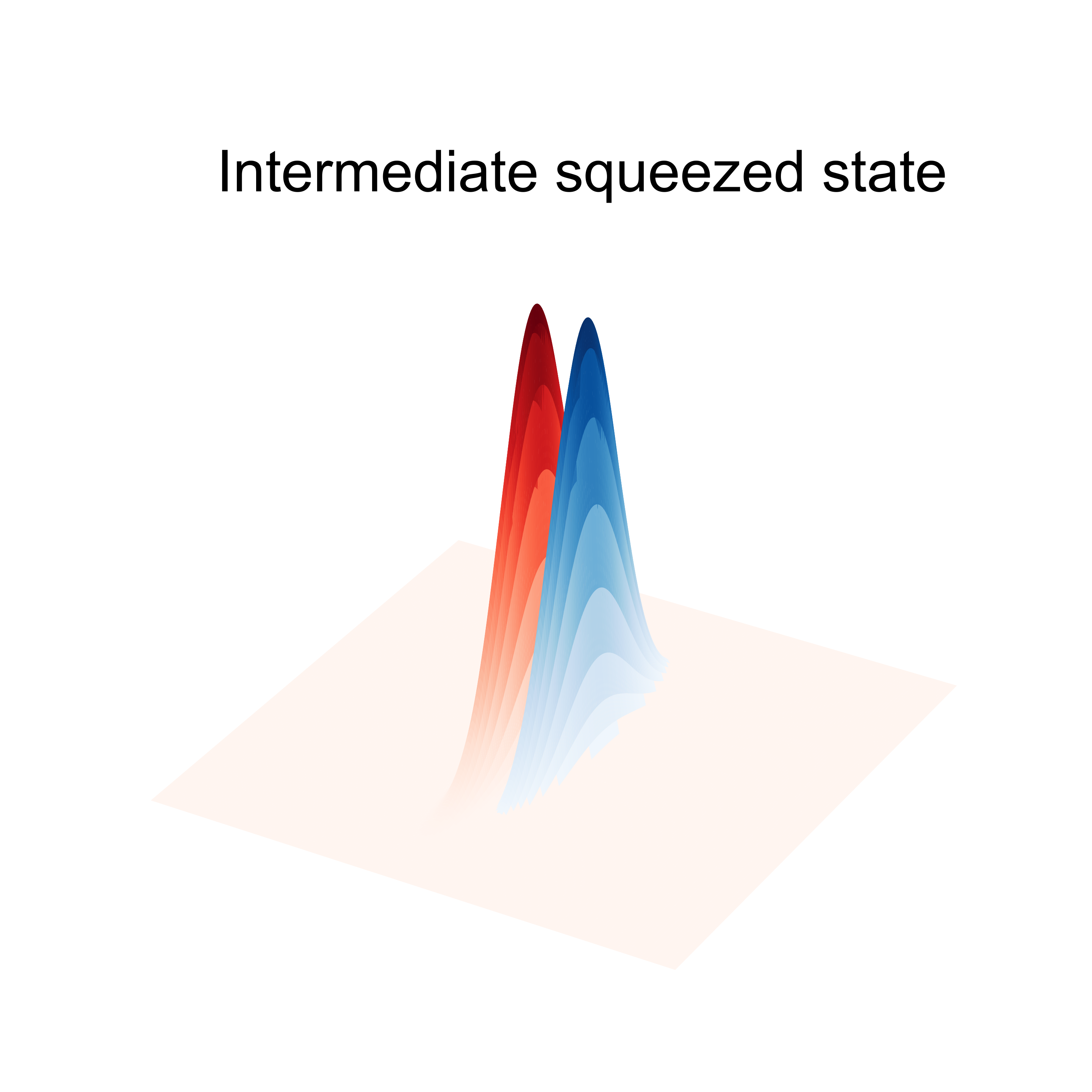}
         \label{fig:wigner2}
         \caption{}
     \end{subfigure}
     \hfill
     \begin{subfigure}[t]{0.3\textwidth}
         \centering
         \includegraphics[width=\textwidth]{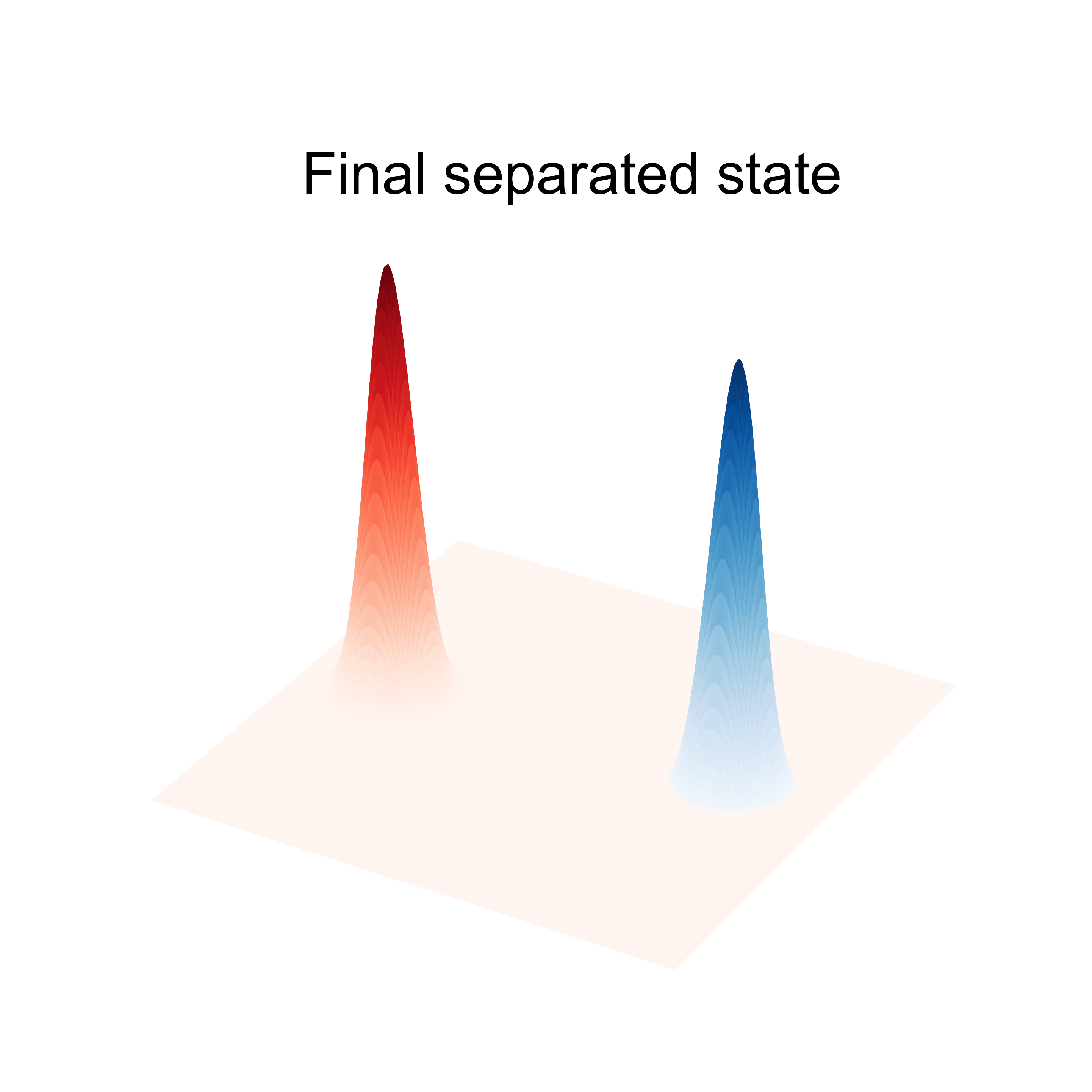}
         \label{fig:wigner3}
         \caption{}
     \end{subfigure}
        \caption{Wigner functions illustrating the splitting and amplification procedure of the two spins. Initially the spin wavefunctions are overlapped, as seen in (a). Maximum squeezing is achieved after half the operation time $t_s/2$, and is represented in (b). In the second half the state is unsqueezed, resulting in amplified displacements. The final superposition of coherent states is shown in (c). The recombination process follows the same steps in reversed order.}
        \label{fig:wigner}
\end{figure*}

and it fully determines the evolution of the squeezing parameter $\zeta$. No analytic closed form exists for the evolution of $\zeta$ so we choose to work with $\tau$ in the rest of this calculation. This avoids having to write separate equations for the squeezing magnitude $r$ and direction $\theta$.

The last term $H_2$ in \eqref{eq:H2gt} can be dealt with by recognizing that the constant term inside the brackets simply leads to the accumulation of a phase:

\begin{equation}
    d\Phi_3 = \harf g(t) dt,
\end{equation}

leaving us to compute the following expression:

\begin{equation}
    \left(1+ig(t)dt\,a^{\dagger}a\right)D(\alpha)S(\zeta)\ket{0}.
\end{equation}
The first step is to pass the number operator to the right of the displacement using the commutation relations:
\begin{equation}
\begin{split}
    (1+ig(t)dta^{\dagger}a) & D(\alpha) = \\ D(\alpha) [1+ig(t)dt\,(a^{\dagger}  & +\alpha^{\ast})(a+\alpha)].
\end{split}
\end{equation}

This factor can be now separated into a phase, a displacement and another factor of the same form as the one on the LHS. Merging the infinitesimal displacement with the displacement operator gives rise to yet another phase. These are given by the following expressions:

\begin{subequations}
\begin{align}
    d\Phi_4 &= g(t)\abs{\alpha}^2dt, \\
    d\alpha_3 &= i\alpha g(t) dt, \\
    d\Phi_5 &= -\frac{i}{2}\left(\alpha d\alpha_3^{\ast}-\alpha^{\ast} d\alpha_3\right).
\end{align}
\end{subequations}

Passing the number operator left through the squeeze operator gives:

\begin{equation}
\begin{split}
    &(1+ig(t)a^{\dagger}a)S(\zeta) = S(\zeta) \\
    &\times [1+ig(t)dt(\hat{a}^{\dagger} \cosh r-e^{-i \theta} \hat{a} \sinh r) \\
    &\times (\hat{a} \cosh r-e^{i \theta} \hat{a}^{\dagger} \sinh r)].
\end{split}
\end{equation}

After acting on the vacuum this turns into an infinitesimal squeeze and a phase:

\begin{subequations}
\begin{align}
    \delta z_2 &= ig(t)dt\, e^{i\theta}\sinh2r, \\
    d\Phi_6 &= g(t)dt\,\sinh^2r.
\end{align}
\end{subequations}

In the end we combine the two squeezes to get another modification of the squeezing parameter and an additional phase:

\begin{subequations}
\begin{align}
    d\tau_2 &= 2ig(t)\tau dt, \\
    d\Phi_7 &= \frac{\cosh^2 r}{4i}(\tau d\tau_2^{\ast}-\tau^{\ast}d\tau_2).
\end{align}
\end{subequations}

Gathering all the terms we get the promised differential equations for $\alpha$ and $\tau$:

\begin{subequations}
\label{eq:motion}
\begin{align}
    \frac{d\alpha}{dt} &= -if(t)e^{i\omega t} - i\alpha^{\ast} g(t) e^{2i\omega t}+i\alpha g(t), \\
    \frac{d\tau}{dt} &= ig(t)\left(e^{i\omega t}+\tau e^{-i\omega t}\right)^2,
\end{align}
\end{subequations}

where we note that the equation for $\tau$ provides an implicit equation for the squeezing $\zeta$ through the definition \eqref{tauzeta}.
We can also express the unitary propagator in a simpler form 

\begin{equation}
\label{unitary}
    U(t) = D(\alpha_0(t)) S(\zeta_0(t)) e^{-i\chi(t)\sigma_3},
\end{equation}

up to a phase. The time functions $\alpha_0(t)$ and $\zeta_0(t)$ are the solutions of equations \eqref{eq:motion} starting from 0. The exponential term arises from the equation \eqref{combS} merging the squeeze operators and has  

\begin{equation}
    \chi(t) = \int g(t) \Re (e^{-2i\omega t} \tau_0 (t)) dt.
\end{equation}

It is easy to see that the displacement $\alpha(t)$ behaves as expected. Consider now the particular example of this type of dynamics given by the SDF+PA driving of the center-of-mass mode. The first term is a result of the SDF and leads to linear increase of amplitude when the force is in resonance with the oscillator. The remaining two terms describe the effect of amplification and lead to exponential increase of amplitude when the RF modulation of the trapping potential is close to twice the frequency of the harmonic oscillator $\omega_i$, as already included in equation \eqref{hamiltPA}. We see that in order to achieve the desired spatial superposition it is sufficient that the SDF function $f(t)$ takes opposite signs for the two spins. This follows because if $\alpha(t)$ is a solution for the pair of driving functions $f(t)$ and $g(t)$ then $-\alpha(t)$ will be a solution of $-f(t)$ and $g(t)$. The two solutions are illustrated in Figure \ref{fig:wigner}.

\section{Method}

We describe here the interference protocol for creating a spatial superposition of the macroscopic particle using the co-trapped atomic ion. We suppose that the ion is initially prepared in the spin state $\vert\uparrow\,\rangle$ and the motional degrees of freedom of the two charged particles are described by thermal density matrices at temperatures $T_i$ and $T_o$ for the in phase and out-of-phase modes respectively. This can be expressed as

\begin{equation}
\begin{split}
    \rho &= \ket{\uparrow}\bra{\uparrow} \otimes \rho^{(i)}_{th} \otimes \rho^{(o)}_{th} \\
    &= \ket{\uparrow}\bra{\uparrow} \otimes \rho_{th}.
    \end{split}
\end{equation}

We start by applying a carrier $\pi/2$ pulse to put the spin degree of freedom in a superposition and get a total density matrix %in block matrix format

\begin{equation}
    \rho = \frac{1}{2} \begin{bmatrix}
\rho_{th} & \rho_{th}\\
\rho_{th} & \rho_{th}
\end{bmatrix}.
\end{equation}

Let us denote by $U_{\uparrow}(t)$ and $U_{\downarrow}(t)$ the evolution operators of the type shown in equation \eqref{unitary}, acting on the two spins. These are the result of acting simultaneously with the SDF \eqref{hami} and the PA \eqref{hamiltPA}, choosing opposite signs of the SDF for the two spins. After applying the interactions driven by $f$ and $g$ for some time the density matrix becomes

\begin{equation}
\label{eq:rho}
    \rho = \frac{1}{2} \begin{bmatrix}
U_{\uparrow}\rho_{th}U_{\uparrow}^{\dagger} & U_{\uparrow} \rho_{th} U_{\downarrow}^{\dagger}\\
U_{\downarrow}\rho_{th}U_{\uparrow}^{\dagger} & U_{\downarrow}\rho_{th}U_{\downarrow}^{\dagger}
\end{bmatrix} = \frac{1}{2} \begin{bmatrix}
\rho_{\uparrow\uparrow} & \rho_{\uparrow\downarrow}\\
\rho_{\downarrow\uparrow} & \rho_{\downarrow\downarrow}
\end{bmatrix}.
\end{equation}

The ideal control sequence will lead to the final state being close to a spatial superposition of the form

\begin{equation}
    \ket{\psi} = \ket{\uparrow}\ket{\alpha_f}_i\ket{0}_o + \ket{\downarrow}\ket{-\alpha_f}_i\ket{0}_o.
\end{equation}

\begin{figure}[t!]
	\includegraphics[width=0.45\textwidth]{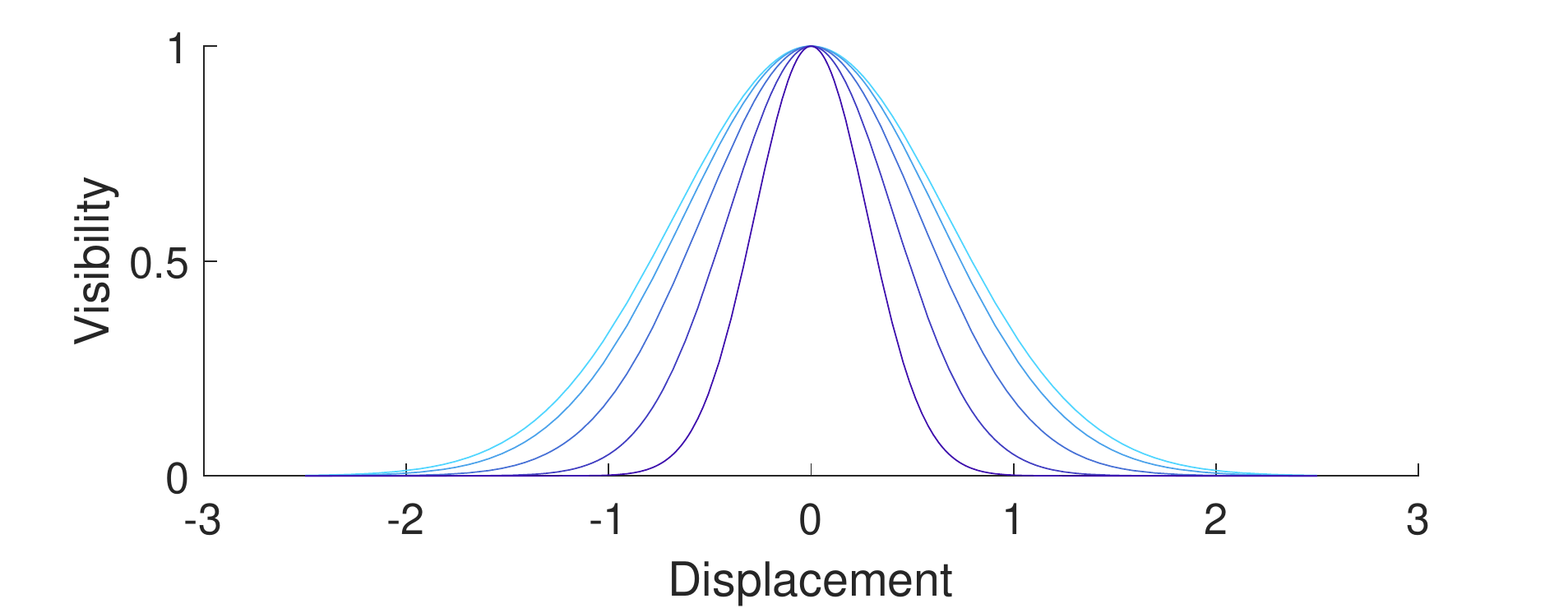}
	\caption{
		Visibility of fringes as a function of displacement between $\uparrow$ and $\downarrow$ spin states. Different colors represent different thermal occupancies of the mode prior to the experiment, from 0.05 (dark blue) to 3 (light blue). Experimental errors may lead to imperfect overlap when the states are recombined and a decrease in visibility. However, the FWHM varies only weakly with temperature, making the experiment robust against faults in cooling the particle to its ground state. Natural quantum harmonic oscillator units are used for the displacement.
	}
	\label{fig:visibility}
\end{figure}

We call the time it takes to split the state into this superposition $t_s$. In order to achieve these states we divide $t_s$ into two equal parts. The SDF is applied consistently throughout both parts, but the phase of the trap potential modulation is switched at the middle point. The effect of this is to squeeze and unsqueeze the mode in order to obtain the desired amplified coherent state at the end, while the squeezing variable $\tau$ performs a full lap and returns to 0. The protocol is illustrated in Figure \ref{fig:wigner}. In order to quantify the benefits of squeezing we introduce the amplification parameter $G = e^r$, where r is related to the maximum value of $\tau$ achieved at $t_s/2$ by equation \eqref{combS}. We can show that the displacement obtained in the presence of squeezing is larger by a factor of $\frac{G-1}{\ln G}$ compared to the value obtained in the same time interval by SDF alone in the validity region of the RWA approximation. For simplicity we only concern ourselves with the sequence described above. Through careful pulse optimisation it may be possible to further speed up the splitting and extend the result outside the validity region of the RWA approximation. We work with the simple expression for the displacement:

\begin{subequations}
\begin{align}
    \alpha_f = \frac{G-1}{\ln G}\alpha_{SDF}, \\
    \alpha_{SDF} = \frac{1}{2}\Omega_R \eta_i t_s,
\end{align}
\end{subequations}

where $\alpha_{SDF}$ is the splitting obtained without amplification.

Proposed theoretical models suggest that this type of superposition state will decohere rapidly once the separation distance $2\alpha_f$ is above a certain threshold. Therefore once we reach the required separation in the state we can stop applying any exterior interactions and let the state evolve according to the harmonic oscillator Hamiltonian alone for a time $t_{int}$, equal to an integer multiple of the in-phase mode period. In order to verify if coherence is retained we invert the splitting operation to recombine the spins and apply a $\mu$-carrier gate described by the operator

\begin{equation}
    U = \frac{1}{\sqrt{2}}\begin{bmatrix}
1 & e^{i\mu}\\
e^{-i\mu} & -1
\end{bmatrix}.
\end{equation}

By measurement of the spin orientation we should be able to observe fringes of intensity when varying the phase $\mu$ \emph{if coherence is retained}. %We find that the
The visibility $\Vis$ of these fringes is related to the trace of the off-diagonal blocks of the density matrix at the end of the evolution by

\begin{equation}
%    \Vis = \abs{\trace{\rho_{\uparrow\downarrow}}},\trace{d}
  \Vis = \abs{\trace{\rho_{\uparrow\downarrow}}},
\end{equation}

where $\rho_{\uparrow\downarrow}$ is defined as in \eqref{eq:rho}. For a better visibility we require the spatial overlap of the states associated with evolution through $U_{\uparrow}$ and $U_{\downarrow}$ to be as high as possible. We observe that if the in-phase and out-of-phase modes are initially not entangled they remain unentangled throughout so that the total visibility of the fringes can be written as a product of the independent visibilities of the in-phase and out-of-phase modes as

\begin{equation}
    \Vis = \abs{\trace{\rho_{\uparrow\downarrow}}^{(i)}}\abs{\trace{\rho_{\uparrow\downarrow}}^{(o)}} = \Vis_i \Vis_o.
\end{equation}

If the unitaries of the type given by equation \eqref{unitary} corresponding to the two different spins $\uparrow$ and $\downarrow$ are different then their action on $\rho_{th}$ is non-trace preserving. %The reduced visibility is then given by an expectation value of the type
we can find the reduction in the visibility from an expectation value of the form

\begin{equation}
    \Vis = \abs{\trace\left\{D(\alpha)S(\zeta)e^{i\Phi \hat{H}}\rho_T\right\}},
\end{equation}

which can be analytically reduced to calculating an overlap of Gaussian functions. This is a simple computational task and we provide code for the calculation of such integrals.

Here the values of parameters $\alpha$, $\zeta$ and $\Phi$ are determined through manipulations of the squeezing and displacement operators. This expression can be evaluated very efficiently using numerical methods. For the case $\zeta, \Phi=0$ there is a simple analytic form

\begin{equation}
\label{eq:visibility}
    \Vis = e^{-\frac{\abs{\alpha}}{2}\coth\frac{\beta \hbar \omega}{2}},
\end{equation}

The visibility is plotted in Figure \ref{fig:visibility} for various thermal occupancy numbers. We can use this expression to compute the visibility of $\rho_{\uparrow\downarrow}$ by writing it as

\begin{equation}
    \Vis = \abs{\trace\left\{D(\alpha_{\uparrow}-\alpha_{\downarrow})\rho_T\right\}}.
\end{equation}

This result proves that the visibility is first order insensitive to both small calibration errors on the scale of one harmonic oscillator length and imperfect cooling to the ground state. The expression can be generalised easily to account for various probabilistic error models (imperfect control of frequency, time of evolution, phase) for which there exists a mapping between the parameter value and the final spin displacement $\alpha_{\uparrow}-\alpha_{\downarrow}$. The case of continuous entropy production is more complex but can be treated phenomenologically as described in section \ref{sec:noise}.

We now perform the experiment for different values of $t_{int}$ and measure the visibility of the fringes each time. This follows an exponential decay law of the type $\Vis \propto \exp(-\Gamma t_{int})$. By fitting the data we can obtain an estimate of the dephasing rate $\Gamma$, and implicitly set an upper bound on the collapse rate. The connection between the two is explored further in section \ref{simulation}.

\begin{figure*}
     \centering
     \begin{subfigure}[t]{0.48\textwidth}
         \includegraphics[width=\textwidth]{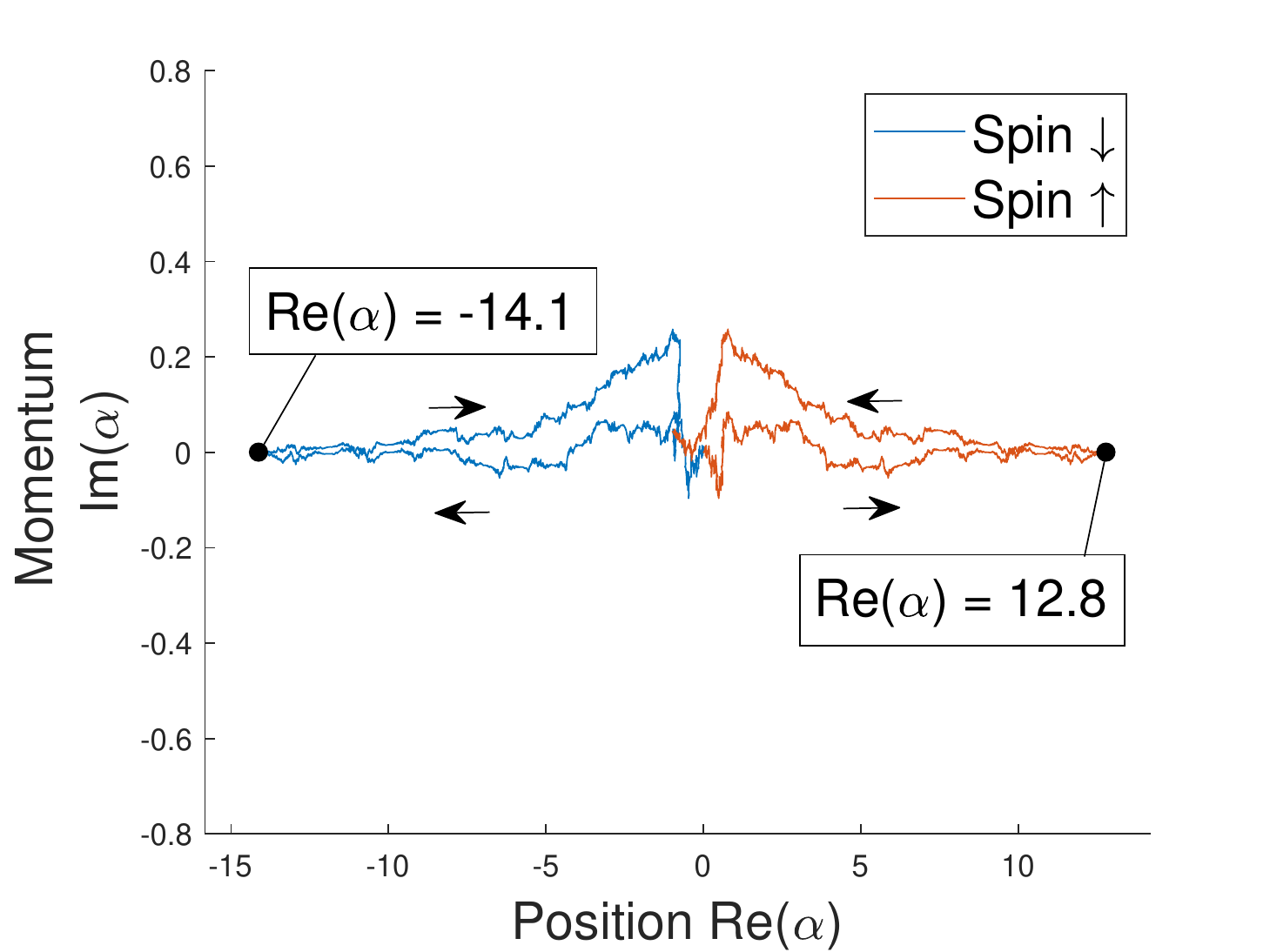}
         \centering
         \subcaption{}
         \label{fig:trajsample}
     \end{subfigure}
     \hfill
     \begin{subfigure}[t]{0.48\textwidth}
         \centering
         \includegraphics[width=\textwidth]{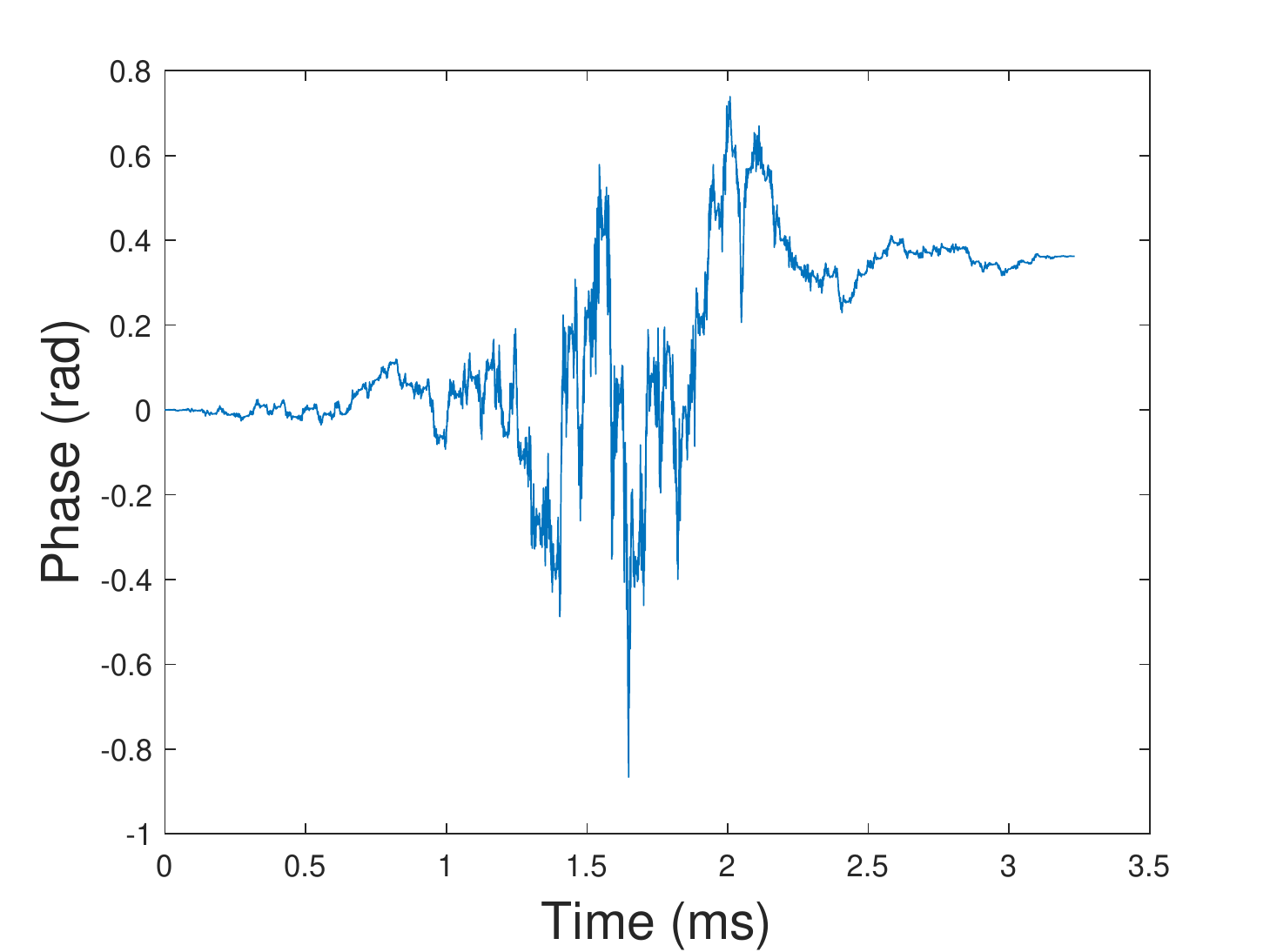}
         \caption{}
         \label{fig:phasesample}
     \end{subfigure}
        \caption{a) Example phase space evolution of the center-of-mass mode for the up (red/light grey) and down (blue/dark grey) spin parts of the superposition in the interaction picture of the harmonic oscillator, in the presence of spontaneous collapses. Natural quantum harmonic oscillator units are used. Central peaks occur at half the splitting time and are due to the amplification of small vertical collapses. Arrows distinguish between the splitting and recombination phases. Real space separation of the superposition of the macroscopic particle is around $\Delta x = 1.2 \hspace{1mm} \mathrm{nm}$. b) Example evolution of the phase difference between the two spins. The variance is sufficiently low to produce fringes when executing runs with $t_{int}=0$.}
        \label{fig:evolsamples}
\end{figure*}

\section{Computer Simulation}
\label{simulation}

We can efficiently perform numerical simulations of equations \eqref{eq:motion} in the RWA regime for the two spins to obtain their trajectories. We also present a simple method to simulate the effect of spontaneous collapses modelled by equation \eqref{modification} directly into the present framework. This has the advantage of retaining independence on the initial conditions, as we are still simulating the evolution of the propagator \eqref{unitary} as opposed to a particular state. The other benefit is the low number of parameters which fully characterise the evolution and do not require explicit numerical manipulations of infinitely dimensional matrices. This is done by performing Monte Carlo simulations of the stochastic equivalent for the von Neumann equation. A detailed description of this procedure is included in Appendix \ref{app:montecarlosimulation}.

Sample trajectories obtained using this method are illustrated in Figure \ref{fig:evolsamples}. In order to obtain these we aimed for a maximum splitting time of $t_s = 1/2\Gamma$, such that the splitting procedure alone does not produce large decreases in visibility. We used an amplification parameter $G = 10$, which appears plausible in the current context, as discussed in Appendix \ref{app:approximations}. In practice we would leave the system to evolve freely for a number of full periods, but this was not included in the simulation. The rotation in phase space due to the natural Hamiltonian of the oscillator does not appear as we are working in the interaction picture. The amplitude of oscillation of the flake is obtained by converting the displacement of the center-of-mass mode to a real distance using the normalizations \eqref{eq:zi} and \eqref{eq:z2} and is around $6$ \AA \space in our simulation. The displacement between the two sides of the superposition which oscillate out of phase is then $\Delta x = 1.2 \hspace{1mm} \mathrm{nm}$ at amplitude. To quantify the behaviour of the spontaneous collapses we chose the GRW \cite{GRW} values for the coherence time $\tau_e = 10^{16}$\,s and a critical length of $\sigma = 10^{-7}$\,m. Since the collapses defined by \eqref{modification} affect both sides of the superposition equally (they do not have any spin or position dependence), the spatial parts of the superposition are perfectly recombined at the end of the evolution, given that no other control errors occur. The central spikes are a result of the squeezing procedure, which amplifies and de-amplifies the small vertical displacements which appear as a result of the collapses. The squeezing parameter is not affected by the kicks, so it also returns to 0 at the end of the evolution. 

The loss of visibility in the final interference fringes arises from the shot-to-shot phase difference between the two spins, as seen in the sample phase trajectory \ref{fig:phasesample}. For this choice of parameters the standard deviation of the final phase difference is close to 1 radian, which should produce a measurable decrease in contrast. We can directly measure the rate at which visibility is lost at maximum amplitude by scanning over a range of values for the time of intermediary free flight. We can approximate this by assuming the final phase follows a Gaussian distribution a mean of zero and standard deviation given by:

\begin{equation}
    \sigma_{\phi} = \alpha_f\frac{z_i}{\sigma}\sqrt{\frac{m}{M}}\sqrt{\gamma t} = \sigma_{0}\sqrt{\gamma t},
\end{equation}

where t is approximately the time of free flight. This follows a diffusive model where random phase kicks of size $\sigma_{0}$ occur at the rate $\gamma$ given by the collapse theory. This would predict a visibility of 

\begin{equation}
    \Vis = \exp(-\sigma_{0}^{2}\gamma t).
\end{equation}

\begin{figure}[t!]
	\centering{}
	\includegraphics[width=0.45\textwidth]{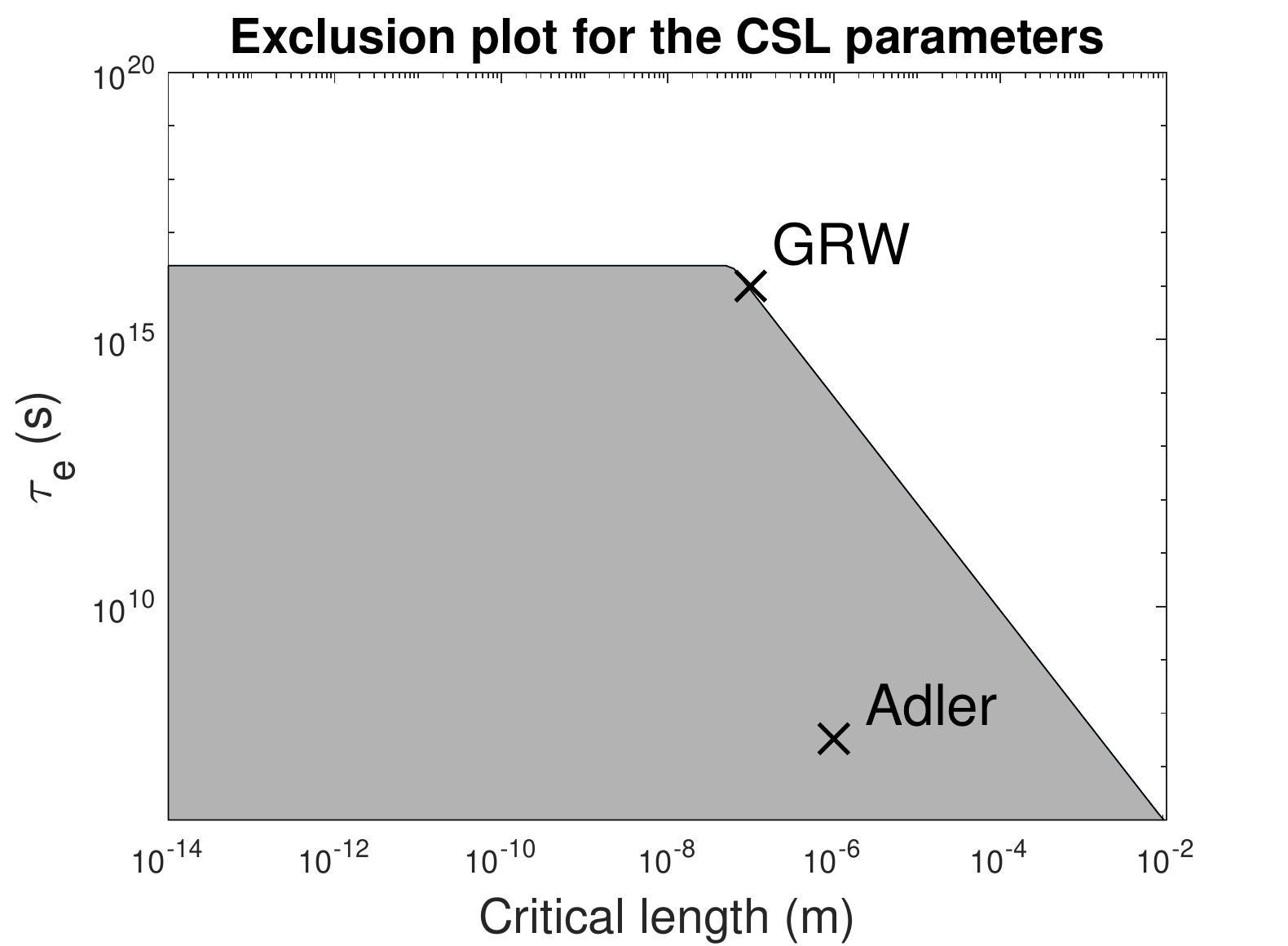}
	\caption{Parameter region excluded by bounding the dephasing rate of spontaneous collapses to $\Gamma = 75 s^{-1}$. The macroscopicity obtained is $\mathcal{M} = 16.4$. The values proposed by GRW \cite{GRW} and Adler \cite{adler2007lower} are indicated.
	}
	\label{fig:exclusion}
\end{figure}

An exponential decay of $\Vis$ with rate $\Gamma = \sigma_{0}^{2}\gamma$. This quantity can be measured experimentally. We follow \cite{Macroscopicity} and parameterize a collapse theory by $\tau_e$ and $\sigma$, where $\tau_e$ is the collapse time of a particle with the mass of an electron and $\sigma$ is the critical length. In general the observed decay rate of the coherences is not solely due to collapses, as discussed in section \ref{sec:noise}, but it allows us to establish an upper bound on the parameter $\tau_e$ for various choices of the critical length $\sigma$. For this relation to hold we require that $\tau_R = 1/\Gamma$ as a result of both collapses and noise-related dephasing must be larger than the splitting time. Assuming that we drive the center-of-mass mode at the largest amplitude allowed by this time constraint and observe fringes which decay in visibility at a rate $\Gamma$, we can plot the parameter exclusion region of this experiment, as shown in Figure \ref{fig:exclusion}.

\section{Error Analysis}\label{sec:noise}

\subsection{Decoherence}

We are especially interested in sources of decoherence which scale with electric charge or mass. In this category we have: stray magnetic fields, current induced in electrodes by Shockley-Ramo effect, Johnson-Nyquist noise in resistors coupled to the electrodes, collision with air molecules, scattering, absorption and emission of thermal photons. Estimation of the strength of these effects is important in deciding if the particle can maintain its coherence throughout the experiment in the absence of spontaneous collapses. In the simplest form we can incorporate these into the evolution through an exponential decay factor $\exp(-\Gamma T)$, where $T \approx 2t_s+t_{int}$ is the total time of the evolution and $\Gamma$ is the decay rate of the off-diagonal terms in position representation. The contribution to this from known effects must be estimated and subtracted from the experimentally observed decay rate of visibility to obtain an upper bound on the collapse rate. A detailed analysis of the heating sources that would affect such a system are described in \cite{steane}. In order to estimate the decay rate with sufficient precision we require that the splitting time alone does not completely destroy the coherence. The duration of the splitting in our simulations is around $t_s \approx 1$\,ms, so it is sufficient to reduce the decoherence rate to the range $\Gamma = 0.1 - 1 \operatorname{ms}^{-1}$. We expect this to be achievable in specialised traps by cooling the electrodes and increasing the dimensions of the trap such as to reduce the effect of fluctuating patch potentials \cite{patchpot}. We estimate the environmental conditions required by this restriction in Appendix \ref{app:decoherence}. Since the magnitude of these effects can be controlled to some extent, we can perform the experiment under multiple conditions and use Richardson extrapolation \cite{extrapolation} to infer the visibility in the no-decoherence limit. This method has recently been explored in the context of error mitigation \cite{mitigationbenjamin,mitigationtemme}.  \\

We neglect errors in the control parameters, since fringe visibility is only affected by differential displacements accumulated on the two paths and numerical simulation shows that some calibration errors cancel out when we consider the displacement of the spins rather than their individual trajectories.

\subsection{Out-of-phase mode}

The trajectory of this mode is governed by off-resonant side-effects of the SDF+PA driving of the in-phase mode. Since the Lamb-Dicke parameters for both of these processes are small and the coupling strength for the out-of-phase mode is of order $\eta_i \eta_o$ we claim that they do not become entangled and can be evolved separately. We note that the oscillating forces of type \eqref{hamo} and \eqref{hamPAoff} generate periodic trajectories, with frequency given by the difference between the driving frequency and the natural frequency of the mode. To maximise our fringe visibility we would like these to return to equilibrium at the end of the experiment, or at least that the spatial overlap of the two spin states is large. To achieve this we choose the evolution times to be integer multiples of the in-phase mode period. This is not a very restrictive constraint as the period is small enough compared to the total evolution time to produce sufficient data but large enough to be accurately measured in practice. We carried out numerical simulations of the out-of-phase mode evolution equations in the form \eqref{eq:motion} to show that this is indeed achieved to a large extent.

\section{Discussion}

This study presents the theoretical foundations for a novel interferometric platform suitable for the study of spatial superpositions of an object that is much more massive than atoms or ions. We outline here the main advantages of the method as well as possible further directions of study.

The core idea behind the design is to avoid having to interact with the macroscopic particle directly. This approach is first order insensitive to the internal degrees of freedom of the object as the oscillation amplitude of around $6$ \AA \space is very small compared to other distances in the trap. As a result we only need to engage the macroscopic observables such as its total charge and center-of-mass position and momenta. In addition, it also enables us to optimize over a broad range of geometries, masses and chemical compositions. The large distance between the two trapped objects leads to very low coupling of the internal degrees of freedom and the center-of-mass mode (i.e. the electric field of the ion is nearly uniform over the flake, so it does not resolve its structure). This opens up the possibility of testing not only for macroscopic objects in their vibrational ground state, but also hot particles \cite{hotinterf}, bringing us a step closer to the realization of true `Schrödinger's cat' type of superpositions. We also show that perfect cooling of the center-of-mass mode to the ground state is not required under the premise that we can perfectly recombine the spins at the end. Laser phase fluctuations in the experiment may affect this in practice, but the results shown in Fig.~\ref{fig:visibility} show that the additional loss of visibility arising from finite temperature is marginal.

We note that there is no fundamental limit to the parameter range that can be tested with this method. The limitations arise strictly from environmental noise and control errors and therefore it can be extended to larger objects as technology improves. The existing  toolbox of ion-trapping techniques can be used straight away with only minor modifications. This is because we only directly interact with the atomic ion, whose behaviour has been widely studied. We also take advantage of the long quantum state coherence times that have been demonstrated for trapped ions. We therefore expect the technique to exclude a wide range of parameter values which appear in attempts to modify the \Schr equation.

The design is very versatile and prototypes would be required to decide on the best trap geometry, atomic ion species and macroscopic particles. Further study is required in order to assess more carefully the impact of different error sources such as black-body radiation, induced currents and mechanical noise on the trap electrodes. Experimental studies are also necessary to establish the difficulty of trapping particles of different charge-to-mass ratios and cooling them to their motional ground state. 

It is also possible to consider different strategies to produce an SDF that does not rely on Raman beams, such as using magnetic field gradients in surface-electrode traps \cite{magnetic}, or microwave radiation \cite{microwave}. This can be accommodated by the current formalism so long as the force is of the type described by equation \eqref{hamiltonian}. A proposed protocol for implementing  double-slit interference of superconducting microspheres by manipulating their quantum states with magnetic fields is directed towards the creation of even larger scale superpositions \cite{RomeroIsart2017} than this paper.

\section{Acknowledgements}

We acknowledge helpful discussions with Andrew Steane, Simone Rijavec and Sebastian Saner in Oxford. We thank the St Peter's College Foundation for financial support.

\begin{appendix}

\section{Experimental considerations for charged particles}
\label{app:charging}

The maximum charge that can be placed on a sphere of radius $R$ in an electric field of magnitude $E$ can be estimated from the Pauthenier limit (as discussed in \cite{Goldwater2019})

\begin{equation} \label{eq:Pauth}
    Q = 4\pi\epsilon_0 R^2 p E,
\end{equation}
where the numerical factor has the value $p= 3$ for conductors. This equation is derived from straightforward electrostatics by considering the superposition of the Coulomb electric field near the surface of uniformly charged shell, $Q/ 4\pi\epsilon_0 R^2$, and the electric field arising from the induced dipole moment. The charge achieved experimentally is generally much less than this theoretical limit but the prediction of proportionality to $R^2$ is a useful guide. The charging of levitated particles is discussed in \cite{Goldwater2019} and the maximum amount of positive charge is much greater than for negative charge.
% \begin{align}
%     \frac{Q}{e} &= 1+\left(\frac{R}{R_0}\right)^2
% \end{align}
%where the constant $R_0=2.2\times 10^{-10}$\,m - $Q/e = 21\times 10^8$. This implies $2\times 10^7$ positive charges for $R= 1\,\mu$m particle can hold up to .
Particles of diameter $5\,\mu$m have been positively charged to $Q/e = 7\times {10}^{6}$ (where $e>0$ is the magnitude of the electron charge)\,\cite{Velyhan2004}. This experimental result is for spherical melamine particles but comparable results should be obtained for discs since the polarisibility and the electric field at the surface depend mainly on dimensions rather than shape. These statements could be made more precise by calculations for a disc as the limiting case of a conducting spheroid as its thickness decreases (at constant radius) but, as indicated above, the electrostatic calculations only give an approximate guide. The following estimate shows the plausibility of obtaining the charge-to-mass ratio required for the scheme proposed in this paper but further experimental  investigation is necessary.\\

We assume an approximately disc-shaped flake of graphene of dimensions $R\simeq 0.8\,\mu$m; this may be a hexagonal platelet that is a natural form for the crystal structure of this material with a single atomic layer. Graphene has an areal density of $7.6\times {10}^{-7}$\,kg\,m$^{-2}$ \,\cite{Heyrovska2016} hence the flake has a mass of $m = 1.7\times {10}^{-18}$\,kg which is $7.1\times {10}^{7}$ carbon atoms. Hence for the micron-sized flake to have a charge-to-mass ratio equal to $7.5\times 10^{-5}$ that of $^{174}$Yb$^+$ requires a positive charge of $6\times 10^{-6}\,e$ per carbon atom corresponding to $Q/e = 430$ to be placed on the flake. There are no obvious physical reasons why this cannot be achieved. The strong covalent bonding ($\sigma$-orbitals) of the hexagonal lattice of carbon atoms will not be significantly affected by removal of some delocalised electrons (from $\pi$-orbitals). The exceptional strength of graphene should also prevent atoms breaking away from the edges of the flake where the electric field will be highest (but less than the fields within atoms). A graphene flake has been loaded into a Paul trap and used for experimental measurements \cite{Kane2010,Nagornykh2017}.   %The estimated electrostatic potential near the flake $V= Q/4\pi\epsilon_0 R = 2\,400$\,V. {\color{red} {$36$\,V} } This is a feasible acceleration voltage for a charged-particle beam with which to bombard the micron-sized graphene flake, e.g.\ He$^+$ accelerated by 5\,kV were used in the experimental work cited above \,\cite{Velyhan2004}. Ion and electron guns up to tens of kV are commercially available but damage to the sample is likely at high voltages as in Focused Ion Beam (FIB) machining.

\section{Approximations}
\label{app:approximations}

The results in the paper rely on various approximation regimes. Understanding these limitations is crucial in deciding whether the approach is experimentally viable when the parameters are optimized. It is also important in the attempt to discover the main difficulties and improve the protocol accordingly. This section acts as a description of the approximation regimes employed in the main text.

\subsection*{Raman transition condition}
From the theory of Raman interaction \cite{home2006entanglement} we have the conditions
\begin{equation}
\begin{split}
    \omega_i \ll \omega_s \ll \Delta, \\
    g \ll \Delta,
\end{split}
\end{equation}
where $\omega_i$ is the frequency of the in-phase mode, $\omega_s$ is the frequency split between the two atomic levels we use to encode spin, $\Delta$ is the frequency offset on the laser driving transitions to the auxiliary state and $g$ is proportional to the laser electric field. If these approximations are satisfied we can obtain the effective SDF Hamiltonian given in equation \eqref{hamiltI}. The effective Rabi frequency is given by:

\begin{equation}
    \Omega_R = \frac{g^2}{2\Delta}.
\end{equation}

Hence the strength of the SDF is limited by the achievable laser power. The (angular) frequency detuning from resonance must be small on the scale of atomic transitions and we expect $\Delta \approx 2\pi\times 10^{11}\,\mathrm{s}^{-1}$ to be a realistic order of magnitude. The value of $g$ must be a few orders of magnitude smaller, e.g.\ for $g \approx 2\pi \times 10^9\,\mathrm{s}^{-1}$ we can achieve Rabi frequencies in the range $2\pi \times 10^6 - 10^7\,\mathrm{s}^{-1}$. Such strong interactions are necessary because of the extremely small Lamb-Dicke parameter of the in-phase mode that we are trying to excite.

\subsection*{Linearisation of Coulomb potential}

We analyse here the assumption that we can safely regard the electrostatic interaction between the two particles as being quadratic for the relevant amplitudes of oscillation. For this we need to take into account that the high degree of squeezing we use to excite the macroscopic particle leads to nonzero values of the wave function at large distances from the equilibrium position. This effect can be estimated by the standard deviation of the position in a squeezed state, which has the well-known expression:

\begin{equation}
    \sigma_x = \sqrt{\frac{\hbar}{2m\omega_i}}\,e^r.
\end{equation}

We denote the amplification factor in our protocol by $G = e^r$. We obtain an upper bound on the squeezing that can be used before the nonlinearity in the electrostatic interaction becomes apparent from :

\begin{equation}
    |z_1-z_2| \approx (b_{1i}-b_{2i})\sigma_x \ll d,
\end{equation}

where d is the distance between the particles at equilibrium. The upper bound on G is then:

\begin{equation}
G \ll \sqrt{\frac{2m\omega_i}{\hbar}}\frac{d}{b_{1i}-b_{2i}} \approx 10^6.
\end{equation}

We can also derive an expression for the allowed displacement size of the out-of-phase mode. If this is quantized by the variable $\abs{\alpha_o}$ we can estimate the physical displacement of the atomic ion by

\begin{equation}
    z_1 \approx \alpha_o b_{1o} \sqrt{\frac{\hbar}{2m\omega_o}},
\end{equation}

and the condition is that this must also be much smaller than d. The resulting constraint on $\alpha_o$ is expressed as:

\begin{equation}
    \alpha_o \ll \sqrt{\frac{2m\omega_o}{\hbar}} \frac{d}{b_{1o}} \approx 10^4.
\end{equation}

\subsection*{Approximate PA Hamiltonian}

In our treatment of the Parametric Amplification protocol we assumed that the Hamiltonian in \eqref{hamiltPA} is the only effect of applying an AC voltage to the endcap electrodes. This is a good approximation as the other terms which appear in the expansion of the interaction are of the same magnitude, but far off-resonant. Other effects include slight shifts in the equilibrium positions of the two particles in the trap, which also has an impact on the natural frequencies of the modes. It may be possible to avoid these unwanted effects by centering the superposed oscillating potential on the macroscopic particle and cancelling the effect on the ion using additional electrodes.

\section{Monte Carlo Simulation}
\label{app:montecarlosimulation}

The stochastic equation for the evolution of a quantum state undergoing a non-unitary process generated by jump operators $D(\alpha)$ \cite{openquantumsys} can be written as:

\begin{equation}
\label{stoch}
\begin{split}
d\ket{\psi(t)}&=-i H dt\ket{\psi(t)}+ \\
&\int d \alpha^2 \left(D(\alpha)-1\right) \ket{\psi(t)} d N(\alpha).
\end{split}
\end{equation}

where $dN(\alpha)$ are independent Poisson increments with rate $\gamma(\alpha) = g(\alpha)/\tau$. (In this Appendix we use the convention $\hbar = 1$.) It can be shown that the von Neumann equation with Lindbladian \eqref{modification} is recovered as the equation of the density matrix corresponding to the ensemble of solutions of \ref{stoch}. It is easily seen that in the case of unitary jump operators the equation remains linear in the state vector. This means we can translate this into an equivalent equation for the propagator, which can be simulated and applied to any initial state, similar to the case of deterministic evolution. This equation reads:

\begin{equation}
\begin{split}
dU(t)&=exp(-iHdt)U(t)+ \\
&\int d \alpha^2 \left(D(\alpha)-1\right) U(t) d N(\alpha).
\end{split}
\end{equation}

A realization of this random process has a simple interpretation. It is a piecewise continuous evolution interrupted in a finite number of positions by jumps of the form $U \mapsto D(\alpha)U$ where $\alpha$ is chosen according to the distribution $g(\alpha)$. The integrated rate of the jumps is then simply given by $\gamma = 1/\tau$. This process is also compatible with the ansatz \eqref{unitary}, due to the simple equation for merging displacements. The localisation effect of the process becomes clear when considering the extra phase introduced by combination of displacements \eqref{combD}. This phase depends on the value of $\alpha$ and it will therefore have a different sign when acting on the two branches of the superposition $U_{\uparrow}$ and $U_{\downarrow}$. To compute the visibility we must average over all realizations of the random process, which amounts to summing over density matrices $\rho_{\uparrow\downarrow}$ which acquired different phases. This means although all realizations return to the initial position with a visibility of 1, the ensemble will have reduced visibility. The algorithm which samples this random process is illustrated in Algorithm \ref{alg}. In the actual implementation we update the parameters in the ansatz of $U$ rather than store $U$ itself (which is infinite dimensional).

\RestyleAlgo{ruled}
\begin{algorithm}[h]
\caption{Monte Carlo Simulation}
\begin{algorithmic}
\STATE $t \leftarrow 0$
\WHILE{$t < t_{f}$}
\STATE $rand \sim$ Uniform$[0,1]$ \COMMENT{check if a kick happens at t}
\IF{$rand > \gamma dt $}
\STATE $U \leftarrow exp(-iHdt)U$
\ELSE
\STATE $\alpha \sim g$
\STATE $U \leftarrow D(\alpha)U$
\ENDIF
\STATE $t \leftarrow t+dt$
\ENDWHILE
\end{algorithmic}
\label{alg}
\end{algorithm}

\section{Decoherence processes}
\label{app:decoherence}

We estimate the decoherence rate of the superposition from some known sources. As for spontaneous collapse, we can describe the strength of these using two parameters, the localization distance $a$ and the value of the decoherence rate $\Gamma$ in the limit $\Delta x \gg a$ of large-distance superpositions. All the formulas have been derived for the simpler case of a sphere of radius R, as opposed to the actual geometry of a graphene flake. However, we expect the difference to be only a numerical factor, which can be considered as being close to unity for these order of magnitude estimates.

\subsection*{Collisions with background gas}

To estimate the pressure needed to keep decoherence rates below the required $\Gamma \simeq 100$\,s$^{-1}$ we consider the rate at which air molecules collide with the macroscopic particle. Following \cite{Rijavec2021} we have:
\begin{equation}\label{eq:aircollision}
    \Gamma = \frac{16 \pi \sqrt{2 \pi}}{3} \frac{\mathrm{P} R^{2}}{\sqrt{m_{\mathrm{air}} k_{\mathrm{B}} \mathrm{T}}},
\end{equation}
where $m_{\mathrm{air}}$ is the average mass of an air molecule. For a particle radius of $R\simeq 0.8\,\mu$m, as above, and room temperature $T=293$\,K, the pressure needed to obtain the decoherence rate stated above is $\mathrm{P}= 1\times 10^{-14} \operatorname{Pa} \equiv 1\times 10^{-12}$\,mbar, which is obtainable using ultra-high vacuum techniques in charged-particle trapping apparatus~\cite{Wineland1998}. Another consideration, also treated in \cite{Rijavec2021}, relates to the thermal de Broglie wavelength of the colliding particles given by 
\begin{equation}
    a = \frac{\pi \hbar}{\sqrt{2 \pi m_{\mathrm{air}} k_{\mathrm{B}} \mathrm{T}}}.
\end{equation}
For $a$ larger than the amplitude of around $6$\,\AA \, in our case, the paths remain indistinguishable and 
decoherence is suppressed. This occurs for temperatures substantially below the threshold value of $70 \operatorname{mK}$. Thus cooling is likely to be technically more difficult than achieving  the required vacuum (but cryo-cooling would help reduce the background gas pressure).

\subsection*{Interaction with thermal photons}

Our charged particle can interact with the background thermal radiation through scattering, absorption and emission of photons. This will carry away which-path information about our system and produce decoherence if the wavelength of the photons is sufficiently small to resolve the superposition separation $\lambda_{th} \lesssim \Delta x$ \cite{decoherence}. We can estimate the temperature necessary for this condition to hold by

\begin{equation}
    \mathrm{T} \gtrsim \frac{\hbar c}{k_{\mathrm{B}}\Delta x} = 1.5\times 10^{6} \operatorname{K}.
\end{equation}

Even if the experiment is performed at room temperature this proves that we can safely neglect the decoherence induced by interactions with all thermal photons.

\subsection*{Shockley-Ramo effect}

The electrostatic coupling of the macroscopic charged particle to the trap electrodes means that oscillations of the charged particle induce currents in the trap's LCR circuit; this is known as the Shockley-Ramo effect \cite{shockleyramo, resistivecooling}. Since the oscillations of the 2 parts of the superpositions are out-of-phase, the state of the macroscopic particle becomes entangled with that of the RF circuit. The resistance in the circuit will lead to effective measurements of the current direction and therefore also introduce decoherence in the state of the entangled macroscopic particle. Since the circuit can be viewed as a quantum harmonic oscillator we can estimate the importance of this effect by looking at the overlap between the 2 states of the circuit corresponding to the 2 parts of the superposition. For this we can use equation \eqref{eq:visibility} and require that the exponent be much smaller than 1. The magnitude of the current induced can be approximated from:

\begin{equation}
    I = \frac{Q}{D}v,
\end{equation}

where $Q$ is the charge on the macroscopic particle, $D$ is the distance between the electrodes and $v$ is the velocity of the particle. At maximum amplitude this is estimated at $I \approx 2\times 10^{-18} \operatorname{A}$. The previously stated condition is then expressed as:

\begin{equation}
    \frac{I}{\sqrt{\frac{\hbar \omega_{LC}}{L}}}\coth{\frac{\beta_{LC}\hbar\omega_{LC}}{2}} \ll 1,
\end{equation}

where L is the effective inductance of the circuit, $\omega_{LC}$ is the frequency and $\beta_{LC}$ is the inverse temperature. For the reasonable choices $\omega_{LC} = 2\pi \times 1 \operatorname{MHz}$, $L = 4 \mu \operatorname{H}$ and $T = 1K$ we get a value of $6.5 \times 10^{-3} \ll 1$.

\end{appendix}

\bibliography{bibliography}% Produces the bibliography via BibTeX.

\end{document}